\documentclass[letterpaper, 10 pt, journal, twoside]{IEEEtran}
\IEEEoverridecommandlockouts                              
\usepackage{amsmath,amsfonts,amssymb,amscd,latexsym,enumerate,bbm,stfloats,theorem}
\usepackage{mathtools} 

\usepackage{graphicx,psfrag}
\usepackage{subfigure}
\usepackage{tikz}
\usepackage{tkz-berge}
\usepackage{float}
\usepackage{cite}
\usepackage{dsfont}
\DeclareMathOperator{\col}{col}
\DeclareMathOperator{\im}{im}

\usepackage{subeqnarray}

\usepackage{epsfig}

 \usepackage{relsize}

\newcommand{\bse}{\begin{subequations}}
\newcommand{\ese}{\end{subequations}}
\newcommand{\bsalign}{
\begin{subequations}
\begin{align}}

\newcommand{\esalign}{
\end{align}
\end{subequations}}
\usepackage[prependcaption,colorinlistoftodos]{todonotes}

\newcommand\oprocendsymbol{\hbox{$\square$}}
\newcommand\oprocend{\relax\ifmmode\else\unskip\hfill\fi\oprocendsymbol}

\newcommand{\bbm}{\begin{bmatrix}}
\newcommand{\ebm}{\end{bmatrix}}

\newcommand{\bsa}{\begin{subeqnarray}}
\newcommand{\esa}{\end{subeqnarray}}

\newcommand{\Xl}{\bar X ^\ell}

\newcommand{\Xpl}{X_\mathsmaller{+}}
{\theorembodyfont{\itshape}\newtheorem{theorem}{Theorem}}
{\theorembodyfont{\itshape}\newtheorem{lemma}[theorem]{Lemma}}
{\theorembodyfont{\itshape}\newtheorem{proposition}[theorem]{Proposition }}
{\theorembodyfont{\itshape}}
{\theorembodyfont{\upshape}}
{\theorembodyfont{\itshape}}
{\theorembodyfont{\upshape}\newtheorem{remark}[theorem]{Remark}}
{\theorembodyfont{\upshape}}
{\theorembodyfont{\upshape}}
{\theorembodyfont{\upshape}}

\DeclareSymbolFont{bbold}{U}{bbold}{m}{n}
\DeclareSymbolFontAlphabet{\mathbbold}{bbold}

\newcommand{\calX}{\ensuremath{\mathcal{X}}}

\newcommand{\calU}{\ensuremath{\mathcal{U}}}

\newcommand{\calV}{\ensuremath{\mathcal{V}}}

\newcommand\Item[1][]{%
  \ifx\relax#1\relax  \item \else \item[#1] \fi
  \abovedisplayskip=0pt\abovedisplayshortskip=0pt~\vspace*{-\baselineskip}}
\usepackage{tikz}
\usepackage{tkz-berge}
\usetikzlibrary{positioning}
\usetikzlibrary{arrows,%
                petri,%
                topaths}%
\usetikzlibrary{decorations.markings}         
\tikzstyle{vertex}=[circle, shading = ball, ball color = white!100!white, minimum size = 15pt, draw, inner sep=0pt]  

\newcommand{\BP}{\noindent{\bf Proof. }}
\newcommand{\EP}{\hspace*{\fill} $\blacksquare$\smallskip\noindent}

\def\be{\begin{equation}}
\def\ee{\end{equation}}
\def\ba{\begin{array}}
\def\ea{\end{array}}

\newcommand{\vs}{\vspace{-2.6mm}}

\newcommand{\R}{\ensuremath{\mathbb R}}

\title{Amidst Data-Driven Model Reduction and Control}

\author{Nima Monshizadeh 
\thanks{Nima Monshizadeh is with the Engineering and Technology Institute, University of Groningen, 9747AG, The Netherlands, {\tt\small n.monshizadeh@rug.nl}}}

\begin{document}
\maketitle
\thispagestyle{empty}
\begin{abstract}
In this note, we explore a middle ground between data-driven model reduction and data-driven control. In particular, we use snapshots collected from the system to  build reduced models that can be expressed in terms of data. We illustrate how the derived family of reduced models can be used for data-driven control of the original system under suitable conditions. Finding a control law that stabilizes certain solutions of the original system as well as the one that reaches any desired state in final time are studied in detail.   
\end{abstract}

\section{Introduction}
Data-driven control is a fertile research venue that is becoming increasingly popular, and has received a big momentum due to the widespread use of data in our infrastructure and technologies.   
The basic idea is to use directly the data generated by the system for analysis and control purposes, as opposed to a system identification followed by a model-based control.

While nonlinear systems are arguably the most relevant dynamics for data-driven applications, they also pose extremely challenging problems. A fundamental difficulty is that a finite set of data may not capture the complexity of nonlinear models.  This is in contrast with linear systems whose behaviours can be fully captured by a finite set of data under suitable conditions \cite{katayama2006subspace,verhaegen2007filtering}. Notably the result of  \cite{willems2005note} on persistently exciting inputs has served as a tool for data-driven control \cite{Markovsky2007,Markovsky2008,Coulson2019,DePersis2019,berberich2019trajectory,Bisoffi2019}.

Under such identifiability conditions, linear models can be replaced by data, and various control problems including stabilization and  LQR can be addressed with data-based LMIs \cite{DePersis2019}. The latter reference also provides data-driven stabilization of nonlinear systems in the sense of the first-order approximation. {For adaptive design of data-based optimal controllers, we refer to  \cite{jiang2012computational, vrabie2009neural}.}
 Model predictive control constitutes another attractive domain for deriving data-driven results under such conditions \cite{salvador2018data,Coulson2019,Berberich2019a}.
In \cite{VanWaarde2019}, by studying the set of all systems compatible with the data, it has been shown that while identifiability is necessary for control problems such as LQR, they can be relaxed in other cases, notably in controllability and stabilization by state feedback.

The difficulty that a finite set of data may not capture the full complexity of nonlinear models, poses an immediate preparatory question; namely, what happens when data is not sufficiently rich to completely explain the behavior of a linear system? In that case, there exists a linear model of a lower order whose complexity matches that of the given data.
The main goal of this note is to uncover such reduced models that can be expressed in terms of data, and subsequently use them for data-driven control.

The connection  of the presented results with those in data-based model reduction stems from the fact that we start off with a finite collection of data from the system rather than an identified state-space model.
In this regard, frequency domain measurements are taken as a tool to construct reduced models in the
Loewner framework \cite{Mayo2007}. In addition, time-domain snapshots are leveraged in \cite{Scarciotti2017}, building on the framework of \cite{Astolfi2010}, to derive reduced models, where notable extensions to the nonlinear case are also provided.

Both in model-based and data-based  model reduction, the eventual reduced models obtained are used for control typically using classical model-based techniques. 
The twist here is that we relate the data-based reduced models directly to the data-driven control problems of the original system.  In that regard, connections to recent results in \cite{DePersis2019}, \cite{VanWaarde2019}, \cite{Berberich2019} are also revealed by studying the reduced models in notable special cases. Moreover, as a consequence of our problem formulation, persistently exciting inputs of the sort  \cite{willems2005note} or \cite{Padoan2017} is not required.

We use the Petrov-Galerkin method \cite{Antoulas:Modred} to project the system onto a space identified by a collection of snapshots from  the state-response. 
In order to write the reduced model in terms of data, 
we apply the projection to a system which is\textit{ isomorphic} to the original model. Given a finite set of data, we discuss how a family of reduced order models can be obtained using the aforementioned method. 
If at least one member of this family satisfies suitable {\em controlled invariance} conditions, then the corresponding reduced model can readily be used as a tool for data-driven analysis/control of the original system.
This has been illustrated on state stabilization as well as finding a control law steering the state of the system to a desired final state; we refer to \cite{baggio2019data} for a different multi-experiments approach on the latter problem.

The structure of the paper is as follows. 
Section \ref{s:input} contains the essential results for the data-driven reduced models.
Connections to data-driven control are drawn in Section \ref{s:data}. Finally, the paper closes with conclusions and future works in Section \ref{s:conc}.

\textbf{Notation.} {The identity matrix of size $n$ is denoted by $I_n$.}
A left inverse of a full column rank matrix $M$ is denoted by $M^\ell$, and a right inverse of a full row rank matrix $N$ by $N^r$. {For a set of vectors $v_1, \ldots, v_k$, we use $\col(v_1, \ldots, v_k)$ to denote in short $[v_1^\top \ldots v_k^\top]^\top.$ The symbol ``$\otimes$" denotes the Kronecker product.}
The rest of the notations are either standard or defined throughout the manuscript.

\section{Data-driven reduced dynamics}\label{s:input}

We consider time-invariant linear systems given by
\be\label{e:sys-input}
x({k+1})=Ax(k)+Bu(k)
\ee
where $x(\cdot)\in \R^n$,  $u(\cdot)\in \R^m$, and $B\in \R^{n\times m}$ has full column rank.
We collect data from the systems in the form of snapshot sequences
\[
X=\bbm x_1 & x_2 & \ldots & x_N \ebm, \quad
U=\bbm u_1 &  u_2 & \ldots & u_N \ebm.
\]
By \eqref{e:sys-input}, the data is mapped to
\be\label{e:yy}
\Xpl:=AX+BU. 
\ee
For a later use, it is useful to define a map 
\[
\sigma(\bar M, \bar N)
:=A \bar M+B \bar N, \quad  \bar M\in \R^{n\times N},  \bar N\in \R^{m\times N}.
\]
Clearly, $\Xpl=\sigma(X,U)$.
We are interested in the case where data do not capture completely the behaviour of the system. 
This can happen in large scale systems when not all the modes of the system are excited by the input, or simply the snapshots are not optimally chosen to allow a complete identification of the system. Another case in point is when part of the data is heavily affected by noise and should be discarded.  A similar situation occurs when data matrices contain small singular values and are practically rank deficient. 



Next, we look for a reduced model that captures the behavior of input-state data in \eqref{e:yy}. To this end, we follow the Petrov-Galerkin projection method \cite{Antoulas:Modred}.
For reasons that will become apparent later, we do not directly apply the projections to \eqref{e:sys-input}, and define first new input variables
 \be\label{e:change}
 v(k):=  u(k) -Kx(k)
 \ee
 for some matrix $K\in \R^{m\times n }$ to be specified later. This yields 
 \begin{align}\label{e:cl}
 x({k+1})&=(A+BK)x(k)+Bv(k).
 \end{align}
 Let $\bar X\in \R^{n\times s}$ be a full column rank matrix satisfying 
 \be\label{e:barX0}
 \im \bar X= \im X:=\calX.
 \ee
 To simplify the presentation, at this point, we choose $\bar X$ as the collection of $s$ independent columns of $X$, with $s=\mathrm{rank}\,(X)$, which is the dimension of $\calX$. We collect the indices of those selected columns in a set  denoted by $\mathcal{I}$. 
 Consistently, we collect $s$ columns of $U$ indexed by $\mathcal{I}$ in a matrix denoted by $\bar U$. Then, we have 
 \be\label{e:E}
 \bbm
 \bar X\\
 \bar U 
\ebm
=
\bbm
X\\
U 
\ebm
E
 \ee
where the matrix $E\in \R^{N\times s}$ is a matrix of zeros and ones selecting the columns indexed by $\mathcal{I}$. 
 
Now, by applying the Petrov-Galerkin projection  $\Pi :=\bar X \Xl$ on \eqref{e:cl}, we obtain the reduced model
 \be\label{e:red-ext}
 \bar x(k+1)= \Xl\, (A+BK)  \bar X\, \bar x(k) + \Xl B v(k),
 \ee
 with $\bar x\in \R^s$, $s\leq n$.
The underlying approximation to obtain \eqref{e:red-ext} is $x(k)\approx \bar X \bar x(k)$, which amounts to projecting the state variables onto the $s$-dimensional (data) subspace $\calX$.
Next, we investigate if the representation in \eqref{e:red-ext} can be expressed in terms of data.
By leveraging \eqref{e:yy} and \eqref{e:E}, we can rewrite \eqref{e:red-ext} as 
\begin{equation}\label{e:red-ext-B}
 \bar x(k+1)= \bar A_B \,  \bar x(k)+ \Xl B v(k), 
\end{equation}
where
\be\label{e:A_B}
\bar A_B=\Xl\,\Xpl E - \Xl B(\bar U-K\bar X).
\ee
For any choice of $K$, \eqref{e:red-ext-B} gives a reduced state-space model in terms of data and the input matrix $B$.

Note that the matrix $B$ appears both in the state matrix as well as in the input matrix of \eqref{e:red-ext-B}.
However, a subtle point is that the expression of the matrix $\bar A_B$ suggests a choice of $K$ that allows us to express $\bar A_B$ in terms of data only, and that is given by
\be\label{e:K-eq}
\bar U=K\bar X.
\ee
Since $\bar X$ has full column rank, the above equation always admits a solution for $K$. In particular, all matrices $K$ satisfying \eqref{e:K-eq} are parametrized by 
\be\label{e:K-exp}
K= \bar U\Xl + R(I_n- \bar X \Xl)
\ee
with $R$ being an arbitrary matrix in $\R^{m\times n}$. 
As a result of this choice, the matrix $\bar A_B$ in \eqref{e:A_B} simplifies to
\be\label{e:barA}
\Xl\,\Xpl E=\Xl\,\sigma(\bar X, \bar U):=\bar A,
\ee
which depends only on data.

On the other hand, in general, the input matrix $\Xl B$ in \eqref{e:red-ext-B} cannot be written in terms of data without identifying the matrix $B$.  Instead, we isolate the data-driven dynamics from the (unknown) input matrix $B$ by defining   
\be \label{e:baru}
\bar u(k):= \Xl\, Bv(k).
\ee

Therefore, we obtain the following dynamics of order $s$:

\begin{equation}\label{e:red-ext-data}
\bar x(k+1)= \bar A  \,  \bar x(k)+  \bar u(k), \quad \bar u\in \bar \calU.
\end{equation}
where $\bar A$ is given by \eqref{e:barA}, and the input constraint set $\bar \calU$ is added since $\bar u$ cannot take arbitrary values in $\R^s$. {This set will be specified later.}  

\vs
\begin{remark}[$\calX=\R^n$]
With a slight abuse of the terminology, we still refer to the notable case of $\calX=\R^n$ and thus $s=n$ as a reduced model. The reason is that even if $\calX=\R^n$, data is not sufficient to obtain system matrices $A$ and $B$. In fact, the latter requires the stronger well-known condition \cite{katayama2006subspace}
\be\label{e:PE}
\im \bbm X \\ U \ebm=\R^{n+m}. 
\ee
\end{remark}

Next, we investigate the conditions under which a concrete relationship between the solution of \eqref{e:cl} and \eqref{e:red-ext-data} can be established. While the representation \eqref{e:sys-input} is isomorphic to \eqref{e:cl}, via \eqref{e:change}, using the latter is much more convenient to draw connections to \eqref{e:red-ext-data}.


We begin the analysis by assuming a controlled invariance condition of the subspace $\calX$, namely \cite{trentelman2012control}
\be\label{e:AB-inv}
A\calX \subseteq \calX + \im B.
\ee

A sufficient condition for the subspace inclusion \eqref{e:AB-inv} in terms of data is provided below.

\vs
\begin{lemma}
The subspace inclusion \eqref{e:AB-inv} holds if
\be\label{e:AB-inv-data}
\im (\Xpl E) \subseteq \calX.
\ee
\end{lemma}
\BP
By \eqref{e:yy} and \eqref{e:E}, we have
\be\label{e:relation-reduced}
\Xpl E= A\bar X+ B \bar U= (A + B K) \bar X
\ee
where $K$ is chosen as in \eqref{e:K-exp}.
Now by \eqref{e:AB-inv-data} and noting that $\im \bar X=\calX$, we obtain
\be\label{e:c-inv}
(A+BK) \calX \subseteq \calX 
\ee
which results in \eqref{e:AB-inv} by \cite[Thm. 4.2]{trentelman2012control}.
\EP

\medskip{}
The reason that \eqref{e:AB-inv-data} is not necessary for \eqref{e:AB-inv} to hold is due to the fact that the matrix $K$ in \eqref{e:c-inv} is restricted to the form \eqref{e:K-exp}, and is not arbitrary. 
Note that if $\calX=\R^n$, then \eqref{e:AB-inv-data} and thus \eqref{e:AB-inv} trivially hold independent of system matrices.

Next to \eqref{e:AB-inv-data}, we need to restrict the set of admissible inputs $v(k)$ and $\bar u(k)$.
Let 
\be\label{e:Vspace}
\calV:= \{v \mid Bv \in \calX\} 
\ee
and 
\be\label{e:Uspace}
\bar  \calU:=\{\bar u \mid \bar X \bar u \in \im B\}
\ee
In order to ensure that $x(k)$ remains in $\calX$, the input $v(k)$ in \eqref{e:cl} must be restricted to the subspace $\calV$.
Then, given $v\in \calV$, we have
\be\label{e:v-u}
Bv(k)=\bar X\bar u(k).
\ee
Moreover, in the opposite direction, obtaining the input $v(k)$ from $\bar u(k)$ is possible providing that $\bar u(k)\in \bar \calU$.

Now, establishing the relationship between the input-state solutions of \eqref{e:cl} and \eqref{e:red-ext-data} is straightforward by noting the {\em model-based} relation
\be\label{e:key}
(A+BK)\bar X= \bar X \bar A.
\ee
The above equality holds due to  \eqref{e:barA}, \eqref{e:AB-inv-data}, and \eqref{e:relation-reduced}.
The relationships between the models are summarized in the following proposition.

\vs
\begin{proposition}\label{p:relation-cl}
Consider the systems \eqref{e:cl} and \eqref{e:red-ext-data}, with $K$ satisfying \eqref{e:K-eq}.
Assume that  \eqref{e:AB-inv-data} holds. 
\begin{enumerate}
	\item Let $x_{[0,T]}$ be a solution to \eqref{e:cl}, with $x(0)\in \calX$ and $v_{[0,T]}\in   \calV$. Then, $\bar x_{[0,T]}= \Xl \, x_{[0,T]}$ is a solution to the reduced order model \eqref{e:red-ext-data} with $\bar u_{[0,T]}=\Xl Bv_{[0,T]} \in 
	\bar\calU$.
	\item Let $\bar x_{[0,T]}$ be a solution to \eqref{e:red-ext-data}, with $\bar u_{[0,T]}\in \bar \calU$. Then, $x_{[0,T]}= \bar X\, \bar x_{[0,T]}$ is a solution to \eqref{e:cl} with $v_{[0,T]}\in \calV$ uniquely obtained from \eqref{e:v-u}.
{	
	\item If the state matrix $A+BK$ in \eqref{e:cl} is stable\footnote{We call a matrix \textit{stable} if all its eigenvalues are inside the unit circle in the complex plane.}, then the state matrix $\bar A$ in \eqref{e:red-ext-data} is also stable.
	
} 
\end{enumerate}
\end{proposition}

\BP
\textit{Item 1)}: From \eqref{e:AB-inv-data}  and thus \eqref{e:c-inv}, it follows that $x(k)\in \calX$ in \eqref{e:cl}, as long as $v(k)$ is restricted to the subspace $\calV$. The result then follows since \eqref{e:red-ext-data} is obtained via projecting the solutions of \eqref{e:cl} onto $\calX$. 

In particular, note that for each $k$, \eqref{e:cl} can be written as
\be\label{e:proof-org-red}
\bar X \bar x(k+1)= (A+BK) \bar X \bar x(k)+ Bv(k)
\ee
where $\bar x$ is uniquely given by $\Xl x(k)$.
Now, by left-multiplying both sides of the above equality with $\Xl$, we obtain 
\[
\bar x(k+1)= \Xl(A+BK) \bar X x(k) \,  \bar x(k)+ \Xl B v(k).
\]
By \eqref{e:key}, the latter simplifies to \eqref{e:red-ext-data}  with $\bar u$ given by \eqref{e:baru}.
Finally, the fact that $\Xl Bv$ belongs to $\bar \calU$ follows from $v\in \calV$.

\textit{Item 2)}: Noting \eqref{e:key}, left-multiplying both sides of \eqref{e:red-ext-data} with $\bar X$ yields \eqref{e:proof-org-red}, with $v$ satisfying \eqref{e:v-u}. Therefore, 
$x_{[0,T]}= \bar X\, \bar x_{[0,T]}$ is a solution to \eqref{e:cl} with $v_{[0,T]}\in \calV$. Uniqueness of $v_{[0,T]}$ is due to the fact that $B$ has full column rank.

\textit{Item 3)}: By \eqref{e:key}, it follows that each eigenvalue of $\bar A$ is an eigenvalue of $A+BK$, which completes the proof.
\EP

\subsection{Family of reduced systems:}

In this subsection, we investigate the family of $s$-order reduced systems, with $s\leq n$, that can be obtained from the approach elaborated before. 

Recall that we chose $\bar X$ as a collection of linearly independent columns of $X$ such that \eqref{e:barX0} is satisfied.
Now, assume that we take a different basis for $\calX$, and denote it by $X'\in \R^{n\times s}$.
In order to mimic the steps that were used to derive the dynamics \eqref{e:red-ext-data}, consistently, we define a matrix $U'\in \R^{m\times s}$ such that 
\be\label{e:F}
\bbm
X'\\
U' 
\ebm
=
\bbm
X\\
U 
\ebm
F.
\ee
Note that the latter is analogous to \eqref{e:E}, where the matrix $F\in \R^{N\times s}$ may not have the special structure of $E$.
Consequently, the state space matrix $\bar A$ in \eqref{e:barA} modifies to
\be\label{e:A'1}
A':= (X')^\ell\,\sigma (X', U')= (X')^\ell \Xpl F.
\ee
Noting that $X'$ has full column rank and $\im X'=\im \bar X$, we have $X'=\bar XS$ for some nonsingular matrix $S$.
Therefore, \eqref{e:A'1} can be rewritten as
\be\label{e:A'2}
A':= S^{-1}\Xl \Xpl F.
\ee

Note that, similar to \eqref{e:K-eq}, the matrix $K$ is chosen such that  $U'=KX'$. Equivalently, the latter can be written as $UF=KXF$ or 
\be\label{e:K-U'}
UF=K\bar X S.
\ee

Next, we observe that the expression of the reduced systems can be simplified by using different coordinates. 
In particular, we have
\[
SA'S^{-1}= \Xl \,\Xpl FS^{-1},
\]
and $\bar X= X FS^{-1}$. By defining $\theta:=FS^{-1}$, $\theta \in \R^{N\times s}$, the latter yields the notable parametrization 
\be\label{e:param}
A_\theta:=\Xl \,\Xpl \theta, \qquad \quad \bar X= X \theta,
\ee
where $A_\theta$ is similar to the matrix $A'$. Noting \eqref{e:K-U'}, the matrix $K$ is chosen such that
\be\label{e:K-theta}
U\theta=K \bar X.
\ee

We record the resulting family of $s$-order reduced models for a later use:
\begin{equation}\label{e:red-ext-data-theta}
\bar x(k+1)= A_\theta  \,  \bar x(k)+  \bar u(k), \quad \bar u\in \bar \calU.
\end{equation}

Note that the terms concerning the input signal remains unchanged since the input matrix $\bar X^\ell B$ in \eqref{e:red-ext-B} first modifies to $S^{-1}\bar X^\ell B$ as a result of the change of the basis from $\bar X$ to $X'=\bar X S$, and then modifies back to $\bar X^\ell B$ due to the change of coordinates used in deriving $A_\theta$.

The treatment preceding \eqref{e:red-ext-data-theta} shows that, modulo similarity transformations, the family of reduced models of order $s$ is provided by \eqref{e:red-ext-data-theta}.

Before concluding this subsection, we note that by comparing \eqref{e:param} and \eqref{e:K-theta} to \eqref{e:barA} and \eqref{e:K-eq}, one can analogously state the result of Proposition \ref{p:relation-cl} with $E$ replaced by $\theta$, and thus $\bar A$ by $A_\theta$.
In that case, the subspace inclusion \eqref{e:AB-inv-data} modifies to
\be\label{e:AB-inv-theta}
\im (\Xpl\theta) \subseteq \calX.
\ee
This is a less restrictive assumption than \eqref{e:AB-inv-data} since $\theta$ can be any matrix satisfying $\bar X=X\theta$.
In fact, in view of \eqref{e:E}, the matrix $\theta$ satisfying the latter equality can be parametrized as
\[
\theta= E+ \tilde{E}, \qquad X\tilde{E}=0.
\]

\vs
\begin{remark}[The controlled invariance assumption]\label{r:inv}
The controlled invariance conditions \eqref{e:AB-inv-data} or \eqref{e:AB-inv-theta} are not required to derive the family of reduced models, but are assumed to guarantee properties of the original model based on the reduced ones.
A straightforward relaxation is given by assuming that a subset $\calX'$ of $\calX$ is controlled invariant, namely \eqref{e:AB-inv} holds with $\calX$ replaced by $\calX'$. Then, we can repeat the steps discussed before using a basis for $\calX'$ rather than $\calX$.  This essentially amounts to discarding part of the data that may spoil the controlled invariance property. 
A challenging open question is  to move from such invariant subspaces to {\em approximately} invariance conditions, and still relates the results to data-driven control of the original system. Such difficulty exists even in model-based reduction since a direct consequence of ``approximation"  is loosing information of the actual system. 
\end{remark}

\section{Towards data-driven control}\label{s:data}

While in model reduction preserving nice properties of the original model is of interest, such as the third item in Proposition  \ref{p:relation-cl}, the opposite direction, namely inferring properties of the original model from the reduced one becomes important in data-driven analysis and control.

\subsection{Stability/stabilization}

After identifying the family of reduced models in the previous section, we are able to look for the models with desired properties. 
The most notable of such properties is stability, which amounts to setting $v(k)=0$ in \eqref{e:cl}. Hence, \eqref{e:cl} simplifies to
\begin{align}\label{e:cl-output}
x({k+1})&=(A+BK)x(k).
\end{align}


\begin{proposition}\label{p:stability-theta}
The following statements hold:
\begin{enumerate}	
	\item Assume that there exists $\theta\in \R^{N\times s}$ with $\bar X= X\theta$\,
	such that 
\eqref{e:AB-inv-theta} is satisfied. Let $K$ be such that \eqref{e:K-theta} holds.
Then any solution $x(\cdot)$ to \eqref{e:cl-output} initialized in $\cal X$ asymptotically converges to the origin if and only if $A_\theta$ given by \eqref{e:param} is stable.

	\item Assume that $\calX=\R^n$, and let $\theta$ be such that $\bar X= X \theta$. Then, $A+BK$ with $K$ satisfying \eqref{e:K-theta} is  similar to $A_\theta$, namely $A+BK= \bar X A_\theta  \bar X^{-1}$.
	In particular, $A+BK$ is stable if and only if $A_\theta$ is stable. 
\end{enumerate}

\end{proposition}

\BP
In the proof below, we use the fact that the result of Proposition \ref{p:relation-cl} analogously holds for the reduced model  \eqref{e:red-ext-data-theta}, where $E$ has been replaced by $\theta$ and $\bar A$ by $A_\theta$. \\
\textit{Item 1, If}: From the first items of Proposition \ref{p:relation-cl}, 
it follows that $\Xl x(\cdot)$ asymptotically converges to zero for any solution to \eqref{e:cl-output} initialized in $\cal X$. By \eqref{e:AB-inv-theta}, $\calX$ is an invariant subspace of \eqref{e:cl-output}, and we conclude that $x(\cdot)$ converges to the origin.  

\textit{Item 1, Only if}: Suppose all solutions $x(\cdot)$ to \eqref{e:cl-output}, initialized in $\cal X$ asymptotically converge to the origin, but $A_\theta$ is \textit{not} stable. Hence, there exists a solution $\bar x(\cdot)$ to \eqref{e:red-ext-data-theta}, with $\bar u=0$, that does not converge to zero. Then, by the second item of Proposition \ref{p:relation-cl}, there exists a solution $x(\cdot)=\bar X \bar x(\cdot)$ to \eqref{e:cl-output} that does not converge to zero either, and we reach a contradiction.

\textit{Item 2}: In this case, $\bar X\in \R^{n\times n}$ is invertible, and we have
\[
A_\theta=\bar X^{-1} \Xpl\theta = \bar X^{-1}(AX+ B U) \theta = \bar X^{-1} (A+BK)\bar X,
\]
where the last equality holds due to \eqref{e:param} and \eqref{e:K-theta}.
\EP


By \eqref{e:AB-inv-theta}, the state-feedback controller $u(k)=Kx(k)$ with $K$ satisfying \eqref{e:K-theta} renders the controlled invariant subspace $\calX$, an invariant subspace of  \eqref{e:cl-output}. Then, Proposition \ref{p:stability-theta} states that the same controller stabilizes the subspace $\calX$ of the original system if and only if the reduced state matrix $A_\theta$ is stable. As data becomes richer, $\calX$ possibly grows and $\calX^\perp$ shrinks. Ultimately, in case $\calX=\R^n$,  stability of $A_\theta$ coincides with that of $A+BK$, and the original closed-loop system is completely stabilized.  

\vs
\begin{remark}\label{r:cvx}
Clearly, $A_\theta$ in \eqref{e:param} is stable if and only if there exists $P>0$ such that
\[
P- A_\theta P A_\theta^\top>0. 
\]
To search for a possible stable reduced model, we need to look for $\theta$ and $P>0$ satisfying the above matrix inequality. Following standard LMI techniques,  an auxiliary variable $Z=\theta P$ can be defined to transform the  above inequality to
\[
\bbm
P & \Xl \Xpl Z\\
(\Xl \Xpl Z)^\top & P
\ebm
>0, \qquad \bar X P - X Z =0.
\] 
which can be efficiently solved using standard LMI packages. Note that the equality constraint is due to $\bar X= X\theta$.  If such $P$ and $Z$ exist,  by 
\eqref{e:K-theta}, the matrix  $K$ can be chosen from $ UZ=K \bar X P$.
\end{remark}
 %
\vs

\vs
\begin{remark}
	In the special case $\calX=\R^n$, the matrix $\theta$ in the second item of Proposition \ref{p:stability-theta} can always be chosen as $X^r \bar X$, where $XX^r=I_n$. 
	This results in 
	$A_\theta= \bar X^{-1} \Xpl X^r \bar X$, which is similar to the matrix $\Xpl X^r$.  The latter coincides with the stabilizability condition in \cite[Thm. 16]{VanWaarde2019}, and can be transformed to the LMI condition in \cite[Thm. 3]{DePersis2019}, in a similar vein as explained in Remark \ref{r:cvx}. The corresponding stabilizing controller is  given by $K=UX^r$. This choice of the controller is at the core of data-driven stabilization results recently reported in \cite{DePersis2019} under the condition of persistently exciting inputs, and thereafter in \cite{VanWaarde2019} and \cite{Berberich2019} without imposing such a condition.   
\end{remark}
\vs
\vs
\begin{remark}[Noisy data]\label{r:noise}
For noisy measurements, the equality in \eqref{e:yy} modifies to
\[
(\Xpl)_\Delta:=AX+BU+\Delta
\]
for some matrix $\Delta\in \R^{n\times N}$. Consequently the reduced state matrix $A_\theta$ in  \eqref{e:param} modifies to
\[
A_\Delta:=A_\theta - \Xl \Delta \theta. 
\]
This means that the family of reduced systems are {\em uncertain} linear systems, with $\Delta$ as the uncertainty block. Note that $\theta$, $A_\theta$, and $\Xl$ are all stated in terms of data. Therefore, the extensive tools from robust control theory can be used to provide robust counterparts of the results presented here \cite{scherer2000linear,scherer1997full,megretski1997system}. The essence of such results would be to restrict the norm of $\Delta$ \cite{DePersis2019}, or more generally to assume that suitable IQCs are satisfied \cite{Berberich2019}.   
Furthermore, since \eqref{e:PE} is not assumed, one can as well opt for discarding parts of data that are heavily affected by noise, such as those associated with fast-varying modes of the system, and follow the presented analysis using the remaining parts of the data.  A completely different approach would be to treat noise as stochastic signals with a given probability distribution and perform a suitable stochastic analysis, see e.g. \cite{Dean2019}.  
\end{remark}

\subsection{Reaching $x_f$ from $x_0$} \label{ss: reaching}

Next, we look into a problem where $v(k)$ in \eqref{e:change} and thus  $\bar u(k)$ in \eqref{e:red-ext-data-theta} may take nonzero values. In particular, we would like to steer the state of the original system \eqref{e:sys-input} from $x_0\in \calX$ to a desired point $x_f\in \calX$ {by using data-based reduced models of the form \eqref{e:red-ext-data-theta}.}

We begin by imposing a  reachability assumption on the data-based reduced model \eqref{e:red-ext-data-theta} rather than on the original system \eqref{e:sys-input}. In particular,  we assume that there exists $W$ with $\im W \subseteq \bar \calU$ such that the pair $(A_\theta, W)$ is reachable, namely
\be\label{e:reach}
\im \underbrace{\bbm W & A_\theta W & \cdots & A_\theta^{\,s-1} W\ebm}_{:=R_{W}} = \R^s,
\ee
for some $\theta$ given by \eqref{e:param}.

Note that we have not used $\im W = \bar \calU$ due to the fact that $\bar\calU$ in \eqref{e:Uspace} generally requires complete knowledge of $\im B$.  
The matrix $W$ instead can be found by using {\em partial} knowledge on the input matrix, namely by using subspaces belonging to $\im B$.  Such partial knowledge may also be obtained directly from data. For instance, by \eqref{e:yy}, we have $\Xpl w\in \im B$ for any $w\in \ker X$.
We remark that the less we know about $\im B$,  the smaller the subspace $\im W$ becomes, and the reachability assumption of the pair $(A_\theta, W)$ thus gets stronger.  

By \eqref{e:reach}, there exists an input $\bar u \in  \im W\subseteq \bar\calU$ that steers the reduced system from any initial state $\bar x_0\in \R^s$ to any final state $\bar x_f\in \R^s$. In particular,
\begin{equation}\label{e:steer-baru}
\bar x_f - A_\theta^s\, \bar x_0= \bbm I_s & A_\theta  & \cdots & A_\theta^{s-1} \ebm  \bar u_{\rm col}
\end{equation}
with $\bar u_{\rm col}:=\col \big(\bar u(s-1), \ldots, \bar u(0)\big)$ and $\bar u(\cdot)\in \im W$.
Then, under the same condition as before, namely \eqref{e:AB-inv-theta}, we can use the mixed open-loop closed-loop control law (see \eqref{e:change})
\be\label{e:input-reach}
u(k)=Kx(k)+ v(k), \quad k=0, 1, \ldots, s-1.
\ee
with
\be\label{e:input-reach-v}
v(k)= B^\ell \bar X \bar u(k),
\ee
and $K$  satisfying \eqref{e:K-theta}, in order to steer the state of the original system from $x_0$ to $x_f$. This is formally stated next.

\begin{proposition}\label{p:reach}
Assume that there exists $\theta\in \R^{N\times s}$ with $\bar X= X\theta$ such that \eqref{e:AB-inv-theta}
is satisfied.
In addition, assume that  there exists $W$ with $\im W \subseteq \bar \calU$ such that the pair $(A_\theta, W)$ is reachable, i.e, 
\eqref{e:reach} holds. Then, the control law \eqref{e:input-reach}, \eqref{e:input-reach-v}, steers 
the system \eqref{e:sys-input} from $x_0\in \calX$ to $x_f\in \calX$, with 
\be\label{e:baru-final}
\bar u_{\rm col}
=(I_s\otimes W)(R_{W})^r (\Xl x_f - A_\theta^{\,s} \Xl x_0),
\ee
$(R_{W})^r$ being a right inverse of $R_{W}$, and $K$ satisfying \eqref{e:K-theta}.
\end{proposition}

\BP
Noting that $x_0, x_f\in \calX$, let $x_0=\bar X \bar x$ and  $x_f=\bar X \bar x_f$ for some $\bar x_0$ and $\bar x_f$.
Since $(A_\theta, W)$ is reachable there exists $\bar u\in \im W$ satisfying \eqref{e:steer-baru}.
We left-multiply both sides of \eqref{e:steer-baru} with $\bar X$, and obtain
\[
\nonumber
\bar X\bar x_f -  \bar X A_\theta^s\, \bar x_0=\bbm \bar X & \bar XA_\theta  & \cdots & \bar X A_\theta^{\,s-1} \ebm 
\bar u _{\rm col}
\]
Observe that by \eqref{e:AB-inv-theta} we have $(A+BK)\bar X= \bar X A_\theta$, similar to  \eqref{e:key}. Bearing in mind that $\bar u \in \im W\subseteq \bar\calU$, we find that 
\begin{multline}
\nonumber
 x_f -  (A+BK)^s\, x_0=\\  \; \bbm B &  (A+BK)B  & \cdots &  (A+BK)^{s-1}B \ebm v_{\rm col} 
\end{multline}
where $v_{\rm col}:= \col(v(s-1), \ldots, v(0))$, and 
$v(\cdot)$ satisfies \eqref{e:v-u} and thus is given by \eqref{e:input-reach-v}.
This means that the input $v$ steers the state of \eqref{e:cl} from $x_0$ to $x_f$. Putting it differently, the input \eqref{e:input-reach} drives the state of the original system \eqref{e:sys-input} from $x_0$ to $x_f$. Therefore, it remains to find the expression for $\bar u_{\rm col}$ in \eqref{e:steer-baru}. Clearly, $\bar u(\cdot)=W \hat u(\cdot)$ for some $\hat u$. By \eqref{e:reach} and \eqref{e:steer-baru},  it is easy to see that $\hat u$ can be chosen as 
\[
{\hat u_{\rm col}}=
(R_{W})^r (\bar x_f - A_\theta^s\, \bar x _0),
\]
where $\hat u_{\rm col}:= \col(\hat u(s-1), \ldots, \hat u(0))$.
Consequently, $\bar u_{\rm col}$ satisfies \eqref{e:baru-final}.
\EP

\vs
\begin{remark}\label{r:mixed}
We note the following three points on the result of Proposition \ref{p:reach}:\\
\textbf{i)} To implement the control input \eqref{e:input-reach}, \eqref{e:input-reach-v}, we need to assume that a left inverse of the matrix $B$ is known. This assumption (partial knowledge) is in general considerably milder than identifying the matrix $B$ itself. 
In particular, knowing a nonsingular $m\times m$ submatrix of  $B$ is sufficient to construct a left inverse $B^\ell$.

\noindent \textbf{ii)} We note that instead of explicitly introducing $W$ in $\bar u(\cdot)$ in \eqref{e:baru-final}, we can directly work with the equation \eqref{e:steer-baru}. In particular, since $(A_\theta, I_s)$ is reachable, this equation has infinitely many solutions for $\bar u_{\rm col}$. Then, it suffices to find one that belongs to $\bar\calU$ or any known subset of $\bar\calU$. Such a solution can replace $\bar u_{\rm col}$ in \eqref{e:baru-final}.
The main purpose of introducing $W$ is to provide an explicit condition on the existence of such a solution, namely $(A_\theta, W)$ must be reachable.  

\noindent \textbf{iii)} For the case $\calX=\R^n$, the assumption \eqref{e:AB-inv-theta} is trivially satisfied. In addition, the matrix $W$ in the proposition can be explicitly written as $W=\bar X^{-1} B^*$ with $\im B^*$ being any known subspace of $\im B$.  For $B^*=B$, the reachability matrix in \eqref{e:reach} associated to the reduced dynamics becomes
\begin{multline*}
\bbm \bar X^{-1} B  & A_\theta \bar X^{-1} B & \cdots & A_\theta^{n-1} \bar X^{-1} B\ebm  \\
=
\bar X^{-1} \bbm B  & (A+BK)B & \cdots & (A+BK)^{n-1}  B\ebm.  
\end{multline*}
Since reachability is invariant under state feedback, the matrix above has full row rank whenever the original pair $(A,B)$ is reachable. The latter is clearly a necessary condition to steer the original system to any desired point; highlighting  again a trade-off between partial knowledge and the strength of the assumption $(A_\theta, W)$ being reachable. Finally note that the proposed control law \eqref{e:input-reach}  differs from the one with minimum control effort \cite{baggio2019data}, particularly due to the presence of the state-dependent term $Kx$.
\end{remark}

\section{Conclusions and future work}\label{s:conc}
In this note, we have explored a middle ground between data-driven model reduction and data-driven control.
A family of reduced models has been obtained by constructing projections based on 
snapshots collected from the state-response of the system, and is expressed in terms of data.
We have observed how the family of reduced models can be used for data-driven control of the original system.
In particular, based on the reduced models, we have shown how to find a control law  that stabilizes an invariant subspace of the system,  as well as a mixed closed-loop open-loop controller that reaches any desired state of the system in final time.

As mentioned in Remark \ref{r:noise}, using noisy data results in uncertain linear systems. 
A detailed treatment of such a case can be carried out using results from robust control and is left for future work. 

Another open question is how to relax the invariance conditions to approximately invariant conditions.
In that case, the reduced models proposed here can still be built, see Remark \ref{r:inv}, however they cannot provide formal stability/performance guarantees for the original system since they no longer represent exact subdynamics of the original model.  Collecting output snapshots rather than state snapshots provides yet another worthy extension, which also relaxes the implicit assumption of knowing the order of the system.

Extension of the results to nonlinear systems would be the ultimate ambition of the preparatory results presented here. Moving from linear geometric control to nonlinear one \cite{nijmeijer1990nonlinear}, and investigating connections to  Koopman invariant subspaces \cite{rowley2009spectral,brunton2016koopman} are among the most promising paths to take. Data-driven reduced models in networked systems \cite{Monshizadeh2014} provides another interesting problem for future research. 

\bibliographystyle{IEEEtran}
\bibliography{Collection}   
\end{document}